\documentclass[showpacs,twocolumn,prb]{revtex4}
\usepackage{amssymb,bbm,bm}
\usepackage{graphicx}
\usepackage{bm}
\usepackage{amsmath}
\usepackage{mathrsfs}
\usepackage[usenames]{color}
\usepackage{epsfig}
\usepackage{hyperref}
\usepackage{subfigure}
\usepackage[sans]{dsfont}
\hypersetup{colorlinks=true, citecolor=blue,
linkcolor=blue,urlcolor=blue }

\begin{document}

\title{Ground-state magnetization of the Ising spin glass: A recursive numerical method and Chen-Ma scaling }

\author{Reza Sepehrinia}\email{sepehrinia@ut.ac.ir}
\affiliation{Department of Physics, University of Tehran, Tehran 14395-547, Iran}
\affiliation{School of Physics, Institute for Research in Fundamental Sciences, IPM, 19395-5531 Tehran, Iran}

\author{Fartash Chalangari}
\affiliation{Department of Physics, University of Tehran, Tehran 14395-547, Iran}

\begin{abstract}
The ground-state properties of quasi-one-dimensional (Q1D) Ising spin glass are investigated using an exact numerical approach and analytical arguments. A set of coupled recursive equations for the ground-state energy are introduced and solved numerically. For various types of coupling distribution, we obtain accurate results for magnetization, particularly in the presence of a weak external magnetic field.  We show that in the weak magnetic field limit, similar to the 1D model, magnetization exhibits a singular power-law behavior with divergent susceptibility. Remarkably, the spectrum of magnetic exponents is markedly different from that of the 1D system even in the case of two coupled chains. The magnetic exponent makes a crossover from being dependent on the distribution function to a constant value independent of distribution. We provide an analytic theory for these observations by extending the Chen-Ma argument to the Q1D case. We derive an analytical formula for the exponent which is in perfect agreement with the numerical results.
\end{abstract}


\maketitle

\section{Introduction}

After several decades of intense study \cite{binder1986spin,fisher1988theory}, spin glasses still remain an active area of research \cite{mydosh2015spin,Stein2011}. In particular, finding their ground state has been a great challenge and developing efficient methods to deal with it is the subject of on-going research \cite{hartmann2011ground,wang2015comparing,liers2003ground}. Even though the absolute zero is not experimentally accessible, knowing the ground state is crucial for a number of reasons: (\textit{i}) because of its own rich structure, exhibiting a phase transition \cite{melchert2009scaling,monthus2014zero} and interesting dynamic properties
\cite{katzgraber2002reversal,katzgraber2007finite}
(\textit{ii}) lower critical dimension of spin glass transition can be determined by studying stiffness exponent of the ground state \cite{amoruso2003scalings,carter2002aspect,hartmann1999scaling,hartmann2001lower} and in low dimensions spin glass phase turns out to be unstable at non-zero temperature (\textit{iii}) in addition to being of fundamental interest, finding the ground state is intimately related to problems in other disciplines like hard combinatorial optimization problems which are currently a challenge in computer science\cite{Lucas2014,zintchenko2015local,hartmann2001optimization}, traveling salesman problem\cite{Lucas2014}, protein folding \cite{trebst2006optimized} etc.

Much of the theoretical work on spin glasses has been based on an Ising model with a random distribution of couplings \cite{binder1986spin,fisher1988theory}. Despite the simplicity of this model a few exact results are presently known \cite{nishimori1980exact,nishimori1981internal,shankar1987exact,shankar1987frustrated,shankar1987nearest} and most results rely on approximations and numerical simulations. In the one-dimensional (1D) model the situation is much simpler and several exact results are known \cite{grinstein1976exact,vilenkin1978random,derrida1978simple,gardner1985zero,ardebili2016ground}. One central property which reveals the structure of the ground state is the magnetic field dependence of magnetization. It is shown \cite{bray1984nonanalytic,chen1982low,gardner1985zero} that the magnetization exhibits a nonanalytic power-law dependence on external magnetic field, $m \propto h^{1/\delta}$. Power-law behavior reflects a scale invariance \cite{gardner1985zero} which is attributed to the zero temperature phase transition in this model. But the scaling exponent turns out to be nonuniversal and depends on details of the distribution function of couplings. We should note, however, that the 1D model lacks frustration which is the fundamental ingredient of spin glass systems \cite{toulouse1977comm}. Zero-temperature magnetization of the two-dimensional (2D) model has also been studied using numerical ground state calculations \cite{mcmillan1984domain,bray1984lower,reger1984investigation,kawashima1992replica,barahona1994ground}. These studies have resulted in the exponent in the range $1/\delta_{\text{2D}} \approx 0.64-0.78$ although the result seemed to be inconsistent with the prediction of droplet picture of spin glasses \cite{fisher1986ordered}. More recently, this issue has been reviewed \cite{liers2007magnetic} and it is suggested that the discrepancy can be removed with larger lattice sizes and corrections to scaling in magnetization. The latter is needed because the magnetic fields that have been used in numerical calculations are not small enough.

In this paper, we study quasi-1D (Q1D) model which is essentially one dimensional and still exhibits nontrivial frustration effects \cite{derrida1978simple}. Moreover, it is still simple enough to allow analytical considerations. First, we introduce an accurate numerical approach to study the ground state of this model in the presence of arbitrary external magnetic field. We generalize the recursive energy method \cite{gardner1985zero,monthus2005spin,monthus2014zero} to the Q1D square lattice. We solve the recursion relations numerically and compute average quantities like energy and magnetization. Through this method, we are able to obtain accurate results at very weak as well as strong magnetic fields. We examine the scaling behavior of magnetization and discuss the dependence of the exponent on the width and the distribution of couplings. Then we present analytical results on scaling in the weak magnetic field which we obtain by extending the Chen-Ma argument \cite{chen1982low} to the Q1D system. We will show that the scaling behavior is strongly modified in the Q1D case even at small widths. Finally, we will summarize and discuss the results.

\section{model and method}

For simplicity, we carry out our discussion for the 2D system, though generalization to higher dimensions is straightforward. The Hamiltonian of the Ising model on a square lattice is given by
\begin{equation}\label{Hamiltonian}
 \mathscr{H}=-\sum_{i,j}(J^h_{ij} \sigma_{i,j} \sigma_{i+1,j}+J^v_{ij} \sigma_{i,j} \sigma_{i,j+1}+h_{ij} \sigma_{i,j}),
\end{equation}
\noindent where the horizontal and vertical couplings $J^h_{ij}, J^v_{ij}$ and fields $h_{ij}$ are uncorrelated random variables taken from given distributions
$\rho(J_{ij})$ and $\rho(h_{ij})$ respectively. As in the 1D chain \cite{gardner1985zero}, the Hamiltonian $\mathscr{H}_{l}$ of a $w\times l$ lattice can be decomposed into the Hamiltonian $\mathscr{H}_{l-1}$ of a $w\times (l-1)$ lattice plus the part which contains the rightmost column
\begin{eqnarray}\label{}
    \mathscr{H}_{l}=\mathscr{H}_{l-1}-\sum_{j} J^h_{l-1,j}\sigma_{l-1,j} \sigma_{l,j}  \hspace{3cm} \nonumber \\ - \sum_{j} J^v_{lj}\sigma_{l,j} \sigma_{l,j+1} -  \sum_{j} h_{lj} \sigma_{l,j}.
\end{eqnarray}
Let us denote by $E^{\alpha}_{l}$ the ground state energy of the lattice with length $l$ for a given configuration of spins in the last column $\{\sigma^{\alpha}_{l,j}, j=1,\cdots, w\}$. Here the superscript $\alpha$ denotes one of the $2^w$ configurations and for each of them, the system has a different ground state. These energies satisfy the following recursion relations
\begin{eqnarray}\label{energy}
E^{\alpha}_{l}=\min_{\beta}\{E^{\alpha}_{l-1}-\sum_{j} J^h_{l-1,j}\sigma^{\beta}_{l-1,j} \sigma^{\alpha}_{l,j}\} \hspace{2cm} \nonumber \\ - \sum_{j} J^v_{lj}\sigma^{\alpha}_{l,j} \sigma^{\alpha}_{l,j+1} -
  \sum_j h_{lj} \sigma^{\alpha}_{l,j}.
\end{eqnarray}
The absolute values of these energies grow linearly with length which means the ground state per spin energy, $E=\lim_{L\rightarrow\infty}E_L^{\alpha}/(wL)$, is finite as we expected from the extensivity of the total energy.

We further simplify Hamiltonian (\ref{Hamiltonian}) by setting uniform magnetic field $h_{ij}=h$ which is then called random-bond model. The ground state of the random-bond model depends on the distribution function of couplings. If couplings are either all ferromagnetic (FM) ($J_{ij}>0$) or antiferromagnetic (AF) ($J_{ij}<0$) the ground state is obviously FM or AF respectively regardless of the distribution function. If zero couplings ($J=0$) are also included in the distribution \cite{jayaprakash1978critical} there would be finite disconnected clusters of spins. Each cluster points up or down independently. Therefore, in the former case, the whole lattice does not support the FM state. By decreasing the density of removed bonds an infinite cluster appears at the bond percolation point which can develop long-ranged order and there would be finite magnetization at the thermodynamic limit. This transition is however purely geometrical and described by the ordinary percolative transition. Far more interesting behavior occurs when the distribution $\rho(J_{ij})$ includes both positive and negative couplings. Then the competition between different interactions leads to frustration which makes the situation more complicated.
If we start with the FM state and increase the density of AF bonds, magnetization decreases and vanishes beyond a critical concentration \cite{grinstein1979ising}. In any case, a nonzero magnetic field will align some clusters and increase the magnetization. In the next section, we will discuss these features in the Q1D lattice by the numerical implementation of Eqs. (\ref{energy}).
\section{Numerical results}
In contrast to what we just discussed, the magnetization of the 1D chain vanishes in zero field regardless of the concentration of FM/AF bonds (see Eq. 5 of Ref. \onlinecite{gardner1985zero}). This can be understood from the fact that even a small concentration of AF couplings would break the chain into clusters of up and down spins and in average there will be no net magnetization. For the same reason, magnetization must vanish also in the Q1D case for zero magnetic field. Unlike the 2D lattice, even a small concentration of AF bonds prevents large FM clusters from percolation. Again the disconnected clusters will have alternating orientations and, therefore, zero average magnetization. In the opposite limit, i.e. the strong magnetic field, all the clusters tend to align and the magnetization saturates.

\begin{figure}[h]
\centerline{\includegraphics[width=8cm]{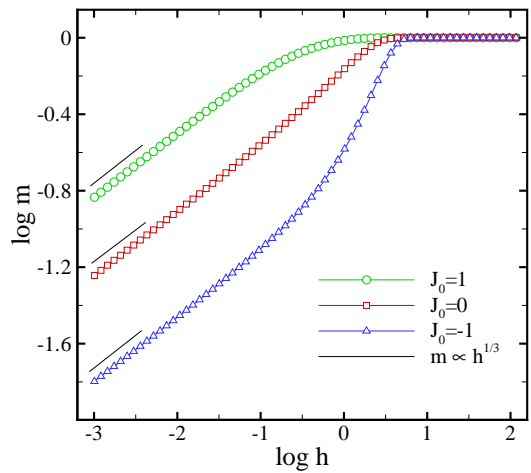}}
\caption{(Color online) Magnetization of single chain ($w=1$) with the Gaussian distribution of couplings as a function of external magnetic field for different values of mean coupling. }
\label{Gauss1}
\end{figure}

\begin{figure}[h]
\centerline{\includegraphics[width=8cm]{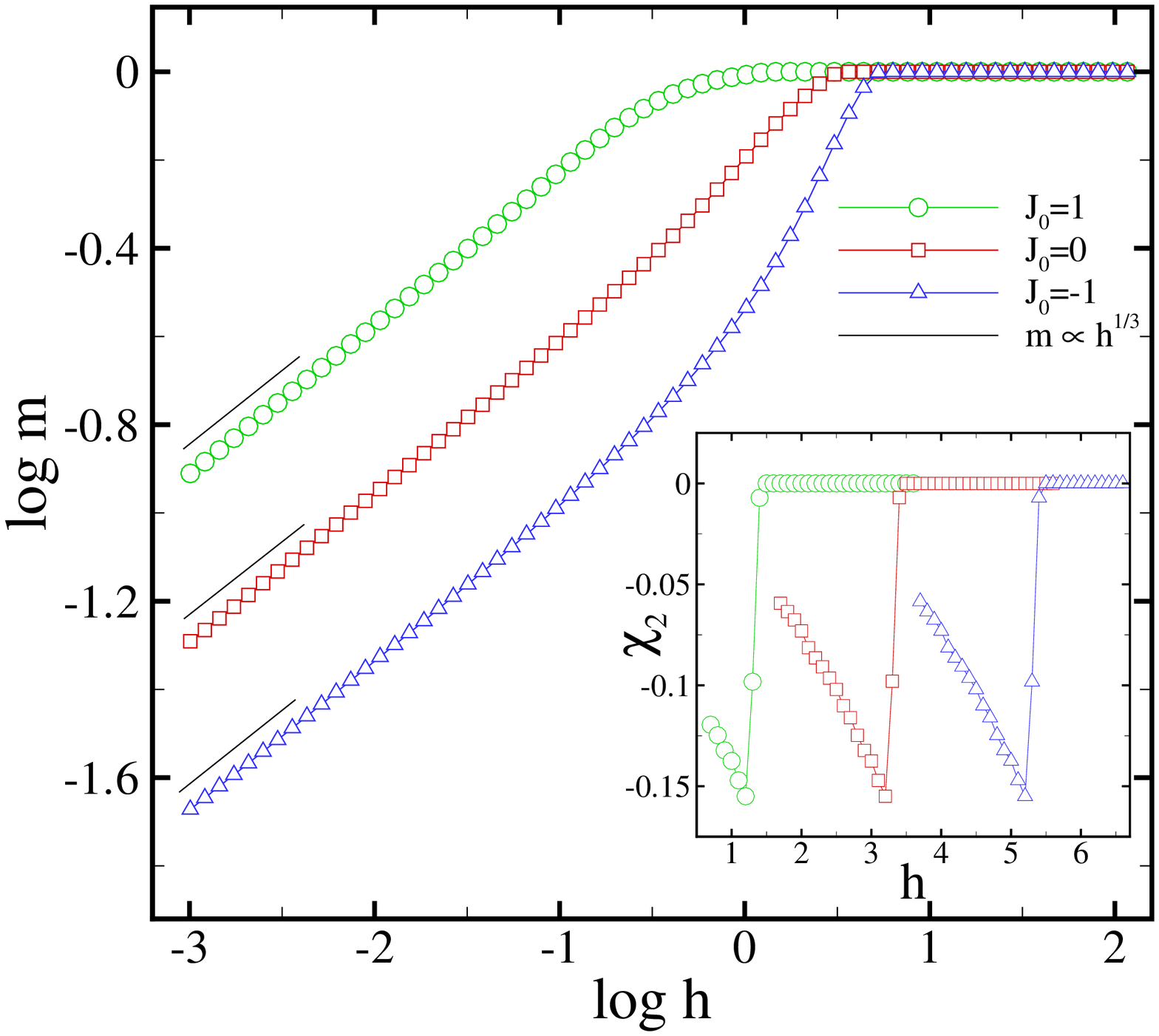}}
\caption{(Color online) Magnetization of single chain ($w=1$) with the uniform distribution of couplings as a function of external magnetic field for different values of mean coupling. Inset: nonlinear susceptibility $\chi_2$ near the saturation point.}
\label{Unif1}
\end{figure}

\begin{figure}[h]
\centerline{\includegraphics[width=8cm]{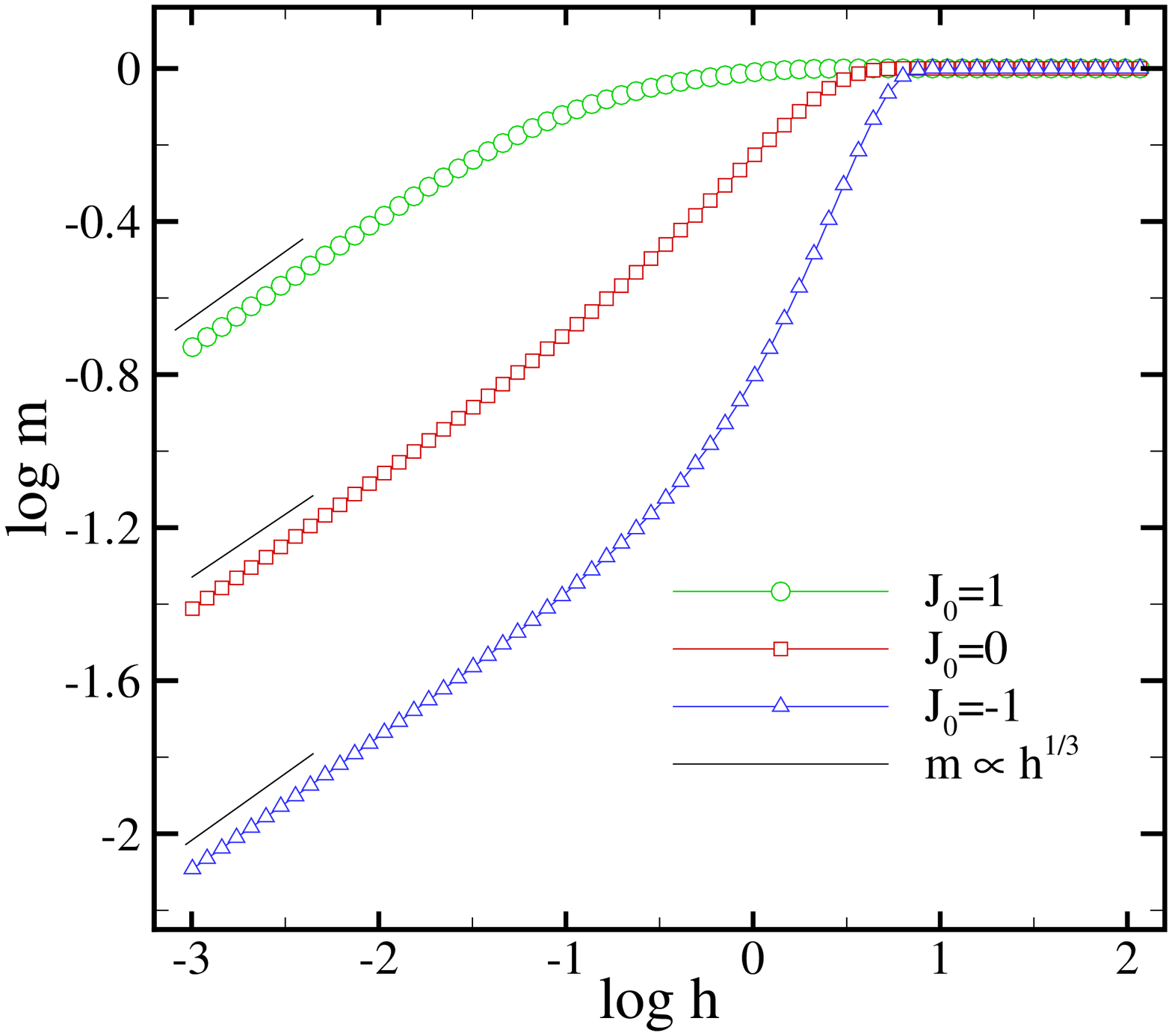}}
\caption{(Color online) Magnetization of ladder ($w=2$) with the Gaussian distribution of couplings as a function of external magnetic field for different values of mean coupling. }
\label{Gauss2}
\end{figure}

\begin{figure}[h]
\centerline{\includegraphics[width=8cm]{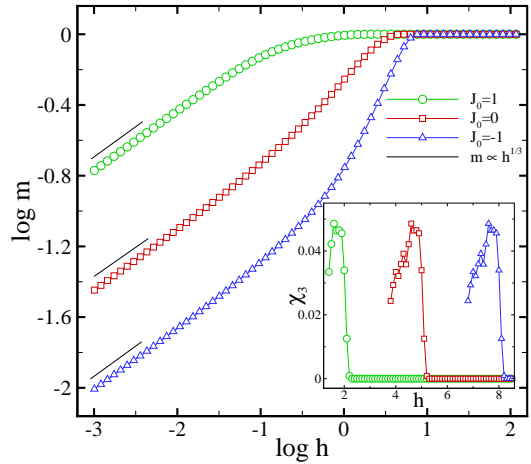}}
\caption{(Color online)  Magnetization of ladder ($w=2$) with the uniform distribution of couplings as a function of external magnetic field for different values of mean coupling. Inset: nonlinear susceptibility $\chi_3$ near the saturation point.}
\label{Unif2}
\end{figure}

Let us first look at the behavior of magnetization in the whole range of magnetic field for different distribution functions $\rho(J_{ij})$ and widths $w$. We use two different distribution functions for this part; Gaussian $\rho(J_{ij})=\frac{1}{\sqrt{2\pi}}e^{-\frac{1}{2}(J_{ij}-J_0)^2}$ and uniform $\rho(J_{ij})=\frac{1}{2J}\Theta(J-|J_{ij}-J_0|)$ distributions with unit variance and mean $J_0$ where $\Theta$ is the Heaviside theta function. Figures \ref{Gauss1}-\ref{Unif2} show the results for chain ($w=1$) and ladder ($w=2$) with different values of mean coupling $J_0$ which correspond to different concentrations of FM/AF couplings. As we expected magnetization vanishes at weak and saturates at strong magnetic field. In all cases, magnetization exhibits power-law dependence with the same power $\frac{1}{3}$ which will be evident later. A notable difference between Gaussian and uniform distributions is in the way magnetization saturates in each case. In the case of uniform distribution, there is a finite value of magnetic field beyond which the magnetization is saturated i.e. the system is fully polarized. This is the point where the magnetic field overcomes all the AF couplings and saturates the magnetization. It is worth noting the differentiability of magnetization at this point. In other words to see whether nonlinear susceptibilities $\chi_n=\partial^n m/\partial h^n$ change continuously or discontinuously. The insets of Figs. \ref{Unif1}, \ref{Unif2} show that the nonlinear susceptibilities $\chi_2$ and $\chi_3$ are discontinuous for chain and ladder respectively. This is of course due to the discontinuity in the uniform distribution function itself.

\begin{figure}[h]
\centerline{\includegraphics[width=8cm]{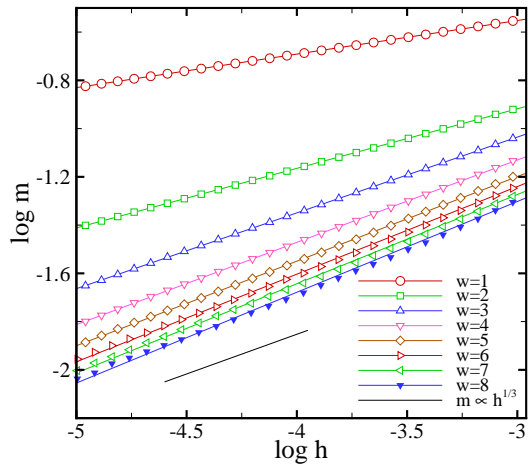}}
\caption{(Color online) Scaling behavior of magnetization in weak external magnetic field for $\mu=-0.7$ and different widths. From linear fitting of data $1/\delta\simeq$ 0.13, 0.24, 0.31, 0.34, 0.35, 0.36, 0.36, 0.37 for $w=1,2,3,4,5,6,7,8$ respectively.}
\label{alpha-0.7}
\end{figure}

\begin{figure}[h]
\centerline{\includegraphics[width=8cm]{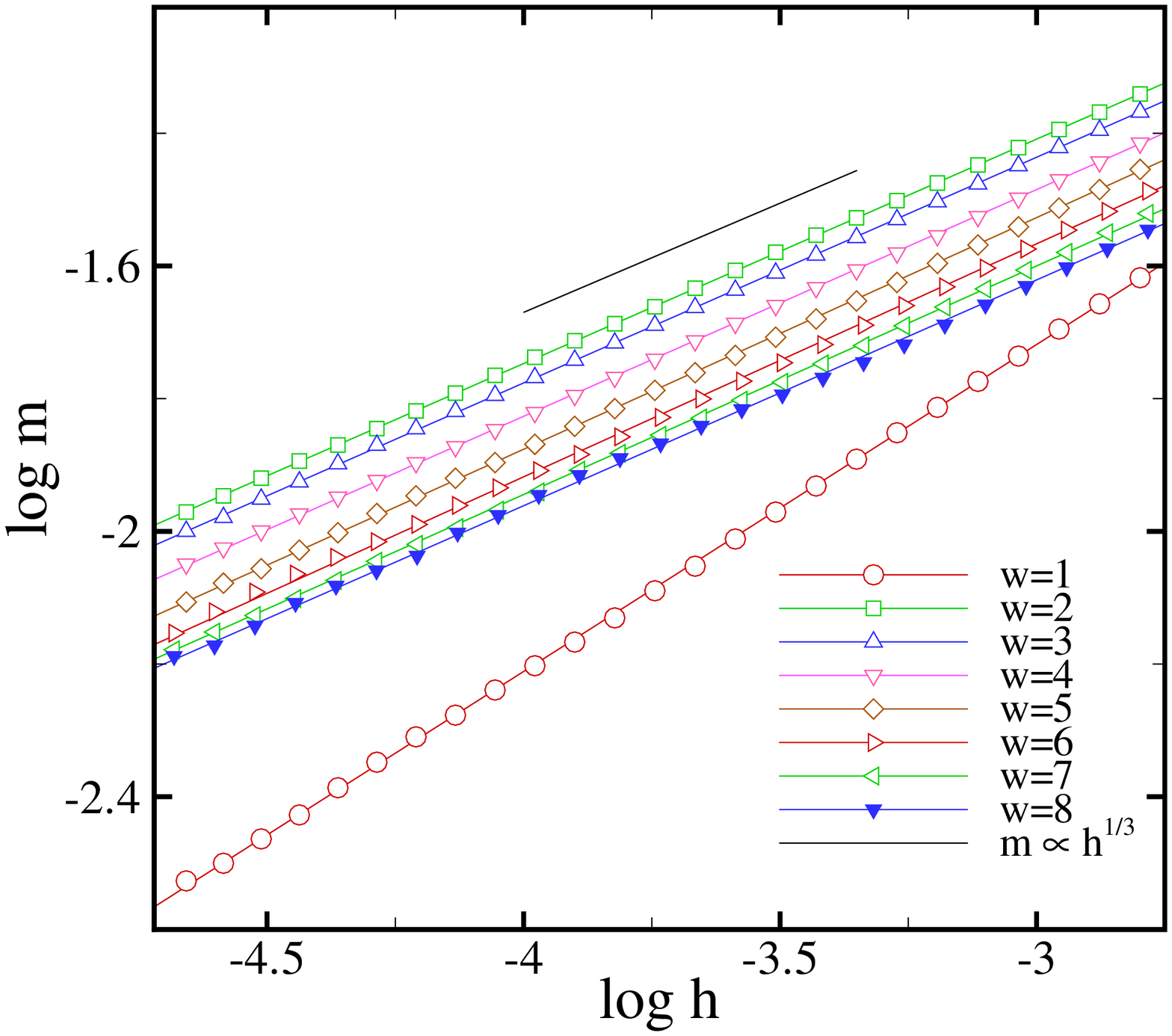}}
\caption{(Color online) Scaling behavior of magnetization in weak external magnetic field for $\mu=1$ and different widths. From linear fitting of data $1/\delta\simeq$ 0.49, 0.33, 0.34, 0.34, 0.34, 0.35, 0.34, 0.34 for $w=1,2,3,4,5,6,7,8$ respectively.}
\label{alpha1}
\end{figure}

\begin{figure}[h]
\centerline{\includegraphics[width=8cm]{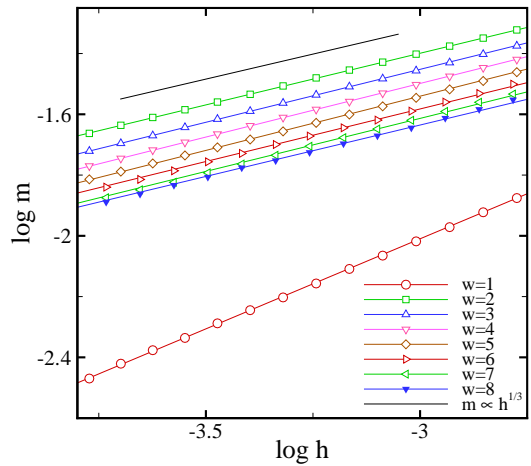}}
\caption{(Color online)  Scaling behavior of magnetization in weak external magnetic field for $\mu=2$ and different widths. From linear fitting of data $1/\delta\simeq$ 0.59, 0.33, 0.34, 0.35, 0.35, 0.34, 0.34, 0.33, for $w=1,2,3,4,5,6,7,8$ respectively.}
\label{alpha2}
\end{figure}

We now focus on scaling behavior of energy and magnetization in weak fields. Here we use the distribution function $\rho(J_{ij})=\frac{1+\mu}{2J^{1+\mu}}|J_{ij}|^{\mu}\Theta(J-|J_{ij}|)$ with different values of exponent $\mu$. The ground state energy  behaves like
\begin{equation}\label{}
  E(h)\simeq E_0+\tfrac{C}{1+1/\delta} h^{1+1/\delta},
\end{equation}
in weak magnetic field, where $E_0=E(0)$ is the energy at zero field and $C$ is a constant. This implies the power-law behavior $m=dE/dh\simeq Ch^{1/\delta}$ for magnetization. Figures \ref{alpha-0.7}-\ref{alpha2} show this scaling behavior
for different widths and different powers $\mu$ of distribution of couplings. The exponent $1/\delta$ is obtained by linear fitting of the data. For $w=1$ the result is in good agreement with the known analytical formula (Eq. 5 of Ref. \onlinecite{gardner1985zero}). For $w>1$ we see two different types of behavior. For $\mu=-0.7$ (Fig. \ref{alpha-0.7}) the exponent depends on $w$ and approaches $\frac{1}{3}$  as $w$ increases. For positive $\mu$ (Figs. \ref{alpha1}, \ref{alpha2}) the exponent is $\frac{1}{3}$ independent of $w$ and $\mu$. In the next section we will provide a theory to explain this scaling behavior.
\section{Analytical results for weak magnetic field: Chen-Ma scaling}
In this section, we would like to provide an explanation for the observed scaling behavior in the weak external magnetic field. We will use the argument which originally has been applied to the 1D model \cite{chen1982low} and is briefly outlined in the following. Since the small magnetic field is only able to overcome weak bonds the behavior will be governed by distribution function in the vicinity of zero coupling which is considered to be like
\begin{equation}\label{rho-power-law}
    \rho(J\rightarrow0)\simeq A|J|^{\mu},
\end{equation}
where the power-law index should satisfy $\mu>-1$ in order to have a normalizable distribution function. Consider domains delimited by weak couplings of the given order $\epsilon$. These domains have average length $\xi\propto \left[\int_{-\epsilon}^{\epsilon} \rho(J) dJ\right]^{-1}\propto \epsilon^{-(\mu+1)}$ and typical net magnetization $m_{\xi}\propto \sqrt{\xi}$ with random direction, up or down. Since the energy cost of flipping all spins in the domain is of order $\epsilon$ (in 1D) and energy gain from magnetic field is $m_{\xi}h$, if the magnetic field is such that $m_{\xi}h \sim \epsilon$, the net magnetization of all these domains will turn in the direction of field therefore the magnetization per spin can be estimated as $m \simeq \frac{N m_{\xi}}{N \xi}\propto\xi^{-1/2}\propto h^{\frac{\mu+1}{\mu+3}}$ where $N$ is the number of such domains. The proportionality constant can not be given by this argument and needs rather detailed calculation which is done in Ref. \onlinecite{gardner1985zero}.

\begin{figure}[t]
\centerline{\includegraphics[width=8cm]{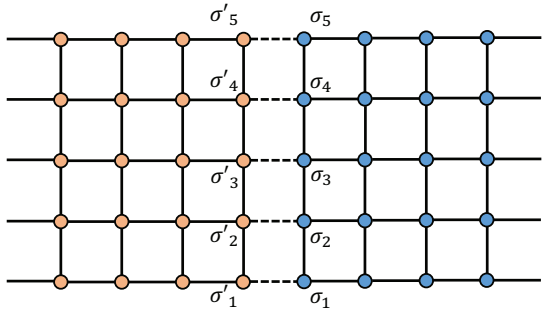}}
\caption{(Color online) Most probable domain wall consisting of weak horizontal bonds shown by dashed lines. }
\label{lattice}
\end{figure}

Now we apply this argument to finite width. The difference is that a domain at its boundaries is coupled horizontally to $w$ spins rather than a single spin. So instead of weak bonds, we should look for a column of weak horizontal bonds in transverse direction (see Fig. \ref{lattice}). To flip the spins of a domain we have to dissatisfy $w$ couplings and the energy required to do this on one side is
\begin{equation}\label{}
  \delta E_w = 2J^h_1 \sigma_{1} \sigma'_{1}+2J^h_2 \sigma_{2} \sigma'_{2}+\cdots+2J^h_w \sigma_{w} \sigma'_{w},
\end{equation}
where $\sigma_i$, $\sigma'_i$ belong to the neighboring domains in the boundary between them.
We now need to have $w$ weak bonds so we modify slightly the derivation for the single chain ($w=1$). Since the couplings are independent, the probability of having $w$ weak bonds is $\left[\int_{-\epsilon}^{\epsilon} \rho(J) dJ\right]^{w}$ so the average size of domains delimited by such boundaries will be $\xi \propto \epsilon^{-w(\mu+1)}$. The similar reasoning that we described above gives
\begin{equation}\label{}
  m\propto h^{\frac{w(\mu+1)}{2+w(\mu+1)}}.
\end{equation}
This is obviously reduced to the previous result for $w=1$. As we can see the exponent depends both on $w$ and $\mu$. However we observed that this is not the case in most circumstances and instead the exponent is constant.

To resolve this contradiction we note that in order to minimize the energy (6) we need not necessarily to have weak bonds. Since we have both positive and negative couplings different terms in (6) can add up to small energy without each term being small in absolute value. We now need to find the density of points where the absolute value of total energy, and not necessarily the individual couplings, is small. For simplicity we consider the case $w=2$, i.e., the ladder structure. Depending on the direction of spins the absolute value of the energy could be $|\delta E_2|=2|J^h_1\pm J^h_2|$. Without loss of generality we consider the plus sign. The probability for this energy to be less than $\epsilon$ is
\begin{equation}\label{}
  P(|\delta E_2|<\epsilon)=\iint\limits_{|J_1+J_2|<\frac{\epsilon}{2}}\rho(J_1)\rho(J_2)dJ_1 dJ_2.
\end{equation}
Since all the coupling are taken from the same distribution we drop the superscripts. By change of variables $x=J_1+J_2$, $y=J_1-J_2$ we have
\begin{eqnarray}\label{}
  P(|\delta E_2|<\epsilon)
  &=& \int_{-\infty}^{\infty}\int_{-\frac{\epsilon}{2}}^{\frac{\epsilon}{2}} \rho\left(\frac{x+y}{2}\right)\rho\left(\frac{x-y}{2}\right)dx dy \nonumber \\
  &\simeq & 2\epsilon  \int_{-\infty}^{\infty} \rho(y)\rho(-y) dy+\mathcal{O}(\epsilon^2).
\end{eqnarray}
So the average distance of such points is of order $\xi \propto \frac{1}{\epsilon}$. As we discussed earlier the magnetic field needed to flip such clusters satisfies $m_{\xi}h\simeq \epsilon$. Using $m_{\xi}\propto \xi^{1/2}$ we have $h\propto \epsilon^{3/2}$. Magnetization per spin would then be $m\sim \xi^{-1/2}\propto \epsilon^{1/2}\propto h^{\frac{1}{3}}$ thus $1/\delta=\frac{1}{3}$.

This is the exponent that we observed in the numerical calculations, however, there is still a problem with Eq. (9). For $-1<\mu<-\frac{1}{2}$ the integral in the first order term diverges which means that the expansion in $\epsilon$ has been incorrect in this range of $\mu$. Let us reconsider the integral in Eq. (8). The distribution function $\rho(J)$ is singular at $J=0$ for this range of $\mu$, so we separate a circle centered at the origin with radius $\frac{\epsilon}{2}$. This part of the integral can be expressed in polar coordinates as $\int_0^{\frac{\epsilon}{2}}\int_0^{2\pi}\rho(r\cos \theta)\rho(r\sin \theta)r dr d\theta$. By substituting from (\ref{rho-power-law}) it can be seen easily that the angular part is convergent and the radial part is proportional to $\epsilon^{2(1+\mu)}$. Note that this is actually the probability of having two weak bonds that we discussed above for general $w$. Non-singular part of the integral is of order $\epsilon$. These two contributions compete with each other and the dominant term is the one with smaller power. Therefore the behavior at small $\epsilon$ is piecewise-defined as a function of power $\mu$. For general $w>1$ the probability is
\begin{equation}\label{}
 P(|\delta E_w|<\epsilon)\propto\left\{
  \begin{array}{cc}
    \epsilon^{w(1+\mu)} & -1<\mu<\frac{1}{w}-1 \\
    \epsilon & \mu\geq\frac{1}{w}-1,
  \end{array}\right.
\end{equation}
and for magnetization we obtain
\begin{equation}\label{m-final}
 m\propto\left\{
  \begin{array}{cc}
    h^{\frac{w(\mu+1)}{2+w(\mu+1)}} & -1<\mu<\frac{1}{w}-1 \\
    h^{\frac{1}{3}} & \mu\geq\frac{1}{w}-1.
  \end{array}\right.
\end{equation}
\begin{figure}[b]
\centerline{\includegraphics[width=8cm]{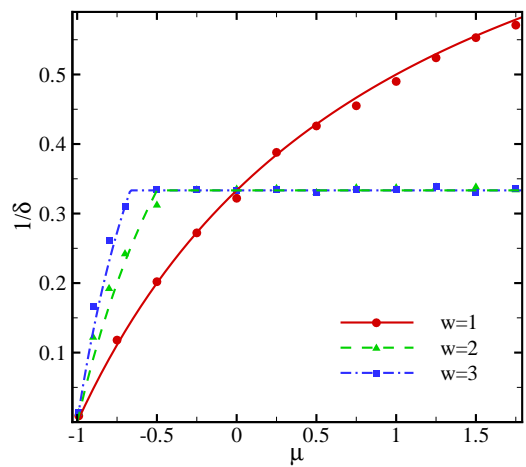}}
\caption{(Color online) Analytical (lines) and numerical (symbols) results for magnetization exponent as a function of the power of distribution function in the vicinity of zero coupling. }
\label{gamma}
\end{figure}

Figure \ref{gamma} shows a plot of magnetization exponent $1/\delta$ as a function of $\mu$ for different values of width $w$. As we can see in this figure, and also in Eq. (\ref{m-final}), for large $w$ the exponent is almost always $1/\delta=\frac{1}{3}$. It is also evident from this result that in Figs. \ref{Gauss1}-\ref{Unif2} the behavior at weak fields must be identical for Gaussian and uniform distributions for which $\mu=0$.

\begin{figure}[t]
\centerline{\includegraphics[width=7cm]{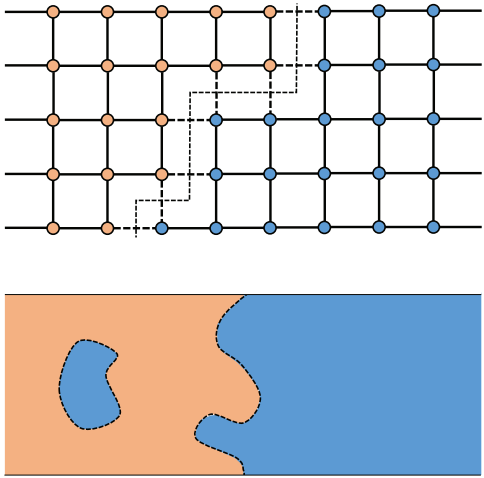}}
\caption{(Color online) General shape of domain walls which is shown by the dashed lines for finite width (top) and continuum limit in large width (bottom).}
\label{domainwall2}
\end{figure}

Further, we would like to point out the contribution of domain walls with a general shape as is shown in Fig. \ref{domainwall2}. The density of such domain walls, in addition to the probability $P(|\delta E_{\ell}|<\epsilon)$, is proportional to a geometrical factor. Here, $\ell$ is the number of bonds between two domains and is proportional to the length of the domain wall. The probability $P(|\delta E_{\ell}|<\epsilon)$ requires the sum of couplings on the domain wall to be small. For large $\ell$ the probability of all couplings being weak is very small and the dominant term will be the analytic term which is proportional to $\epsilon$. This can be seen from the fact that, according to the central limit theorem, the sum of large number of random variables is normally distributed so $P(\delta E_{\ell})\propto \exp(-c_{\ell}\delta E_{\ell}^2)$ where $c_{\ell}$ is a constant. Therefore, $P(|\delta E_{\ell}|<\epsilon)\propto \epsilon$ which leads to the same exponent. However, as the width increases the geometrical factor and also domains with closed boundary (Fig. \ref{domainwall2}, bottom) become important and could play a role in restoring the 2D result.

\section{Summary and discussion}
We have investigated the ground state of the Ising spin glass with a particular emphasis on scaling properties. We have introduced an accurate numerical procedure to calculate the thermodynamic quantities without explicitly obtaining the spin configuration of the ground state. We have studied ground state energy and magnetization for different widths and distribution of couplings as a function of external magnetic field and concentration of FM/AF bonds. We found that power-law behavior in the weak magnetic field is maintained in Q1D geometry under various conditions. Even at the small widths, we see large deviations from the 1D result. In marked contrast to the 1D model, the scaling exponent becomes independent of width and distribution function when the power-law index of distribution function near zero coupling, $\mu$, exceeds a certain value. In the second part of the paper, which is devoted to a theoretical explanation of the observed scaling behavior, we derive an analytical formula for the exponent using the Chen-Ma argument. The additional possibility of constructing weak interfaces in Q1D geometry modifies the scaling behavior and interprets the observed constant exponent. For some range of the exponent $\mu$, which depends on the width, the magnetization exponent is a function of parameters, i.e., is nonuniversal.  By increasing the width this nonuniversal region gets smaller, therefore our results suggest a universal exponent, $1/\delta=\frac{1}{3}$, for higher widths. However, the existing results \cite{mcmillan1984domain,bray1984lower,reger1984investigation,kawashima1992replica,barahona1994ground} for the 2D exponent are approximately in the range $1/\delta_{\text{2D}}\approx 0.64-0.78$ which is different from our extrapolated result. The reason for the difference could be the fact that (\textit{i}) the system which we considered is basically Q1D or (\textit{ii}) more complex domain walls (Fig. \ref{domainwall2}) are not taken into account in our analysis and the widths in our numerical work are not large enough to capture their contribution. The numerical approach that we applied here to the Q1D system can be combined with finite size scaling analysis to study the phase transition that occurs in the ground state of the 2D spin glass which also will be considered in future work.

\section{aknowledgement}

The authors would like to acknowledge financial support from the research council of University of Tehran for this research.

\bibliography{myref}
\bibliographystyle{apsrev}

\end{document}